\documentclass[pdflatex,sn-mathphys-num]{sn-jnl}

\usepackage{graphicx}%
\usepackage{multirow}%
\usepackage{amsmath,amssymb,amsfonts}%
\usepackage{amsthm}%
\usepackage{mathrsfs}%
\usepackage[title]{appendix}%
\usepackage{xcolor}%
\usepackage{textcomp}%
\usepackage{manyfoot}%
\usepackage{booktabs}%
\usepackage{algorithm}%
\usepackage{algorithmicx}%
\usepackage{algpseudocode}%
\usepackage{listings}%
\usepackage{lmodern}
\usepackage{fix-cm}
\usepackage[T1]{fontenc} %
\usepackage{anyfontsize}

\newtheoremstyle{thmstyleone}
  {3pt}      
  {3pt}      
  {\itshape} 
  {}         
  {\bfseries}
  {.}        
  { }        
  {\thmname{#1}\thmnumber{ #2}\thmnote{ (#3)}}

\theoremstyle{thmstyleone}
\newtheorem{theorem}{Theorem}[section]

\begin{document}

\title[Stability analysis of geodesics in dynamical Chern-Simons black holes: geometrical perspective]{Stability analysis of geodesics in dynamical Chern-Simons black holes: a geometrical perspective}

\author[1]{\fnm{Tonatiuh} \sur{Tiscareño}}
\email{tonatiuh.tiscareno@fisica.uaz.edu.mx}

\author[1]{\fnm{Benito } \sur{Rodríguez}}
\email{benito.rodriguez@fisica.uaz.edu.mx}

\author[1]{\fnm{Javier} \sur{Chagoya}}
\email{javier.chagoya@fisica.uaz.edu.mx}

\affil*[1]{\orgdiv{Unidad Académica de Física}, \orgname{Universidad autónoma de Zacatecas}, \orgaddress{\street{Calzada Solidaridad esquina con Paseo a la Bufa S/N}, \postcode{98060}, \state{Zacatecas}, \country{México}}}

\abstract{We apply the Kosambi-Cartan-Chern theory to perform an extensive examination of Jacobi stability of 
geodesics around rotating black hole solutions to dynamical Chern-Simons gravity, a theory that introduces modifications to General Relativity via a scalar field non-minimally coupled to curvature scalars. We present a comparative study 
between Jacobi and Liapunov stability, pointing out the advantages of the more 
geometrical method over the usual Liapunov approach.}


\maketitle

\section{Introduction}\label{sec1}

This study focuses on timelike geodesics around black holes in the context of modifications to General Relativity (GR) that are introduced through a scalar field that is non-minimally coupled to curvature scalars~\cite{Jackiw_2003, Yunes_2011}. Specifically, we work with dynamical Chern-Simons (dCS) gravity \cite{Smith:2007jm}, a notable variant of these theories, which has been extensively explored due to its deviations from GR in rotating black hole solutions.

Although spherically symmetric spacetimes remain unchanged in dCS gravity, axially symmetric solutions deviate from those in GR. The groundbreaking work by Yunes and Pretorius \cite{Yunes_2009} presented the first rotating black hole solution under the slow rotation and small coupling approximations in dCS gravity. This solution introduces parametric modifications to the Kerr metric, resulting in corrections to the positions of photon orbits due to the presence of a scalar field or ``scalar hair'' \cite{Grumiller_2008,Alexander2009}. Additionally, the dCS solution has the potential to result in violations of the Hawking-Penrose theorem~\cite{Alexander_2021}.

To address the complexity of the geodesic equations in dCS spacetime, we begin by studying linear stability (Lyapunov) and phase portraits to investigate the behavior of the dynamical system. However; due to the nature of the system, nonlinear stability methods are essential for a more comprehensive understanding of its global behavior, therefore we complement the linear Lyapunov analysis with the non-linear Kosambi-Cartan-Chern (KCC)~\footnote{Sometimes also known as Jacobi stability.} theory for a thorough analysis.

 The KCC theory, developed by Kosambi, Cartan, and Chern~\cite{Chern,Kosambi1933,Cartan1933}, is a robust framework for studying the non-linear stability of dynamical systems of higher dimension and even non-autonomous systems; it has multiple applications in gravitation and cosmology~\cite{Cardoso_2009,boemer2010,aceña2019circulargeodesicsstabilitystatic}. Furthermore, in~\cite{abolghasem} a comparative study between Jacobi stability, a core aspect of KCC theory, and Lyapunov stability is presented, identifying cases where the two approaches yielded differing results. A similar study has been conducted for black hole solutions in GR, specifically for Kerr and Schwarzschild, metrics~\cite{Hossein,Singh_2022} and other related theories like New masive gravity and traversable wormhole spacetime ~\cite{galaxies8010014,Giri2022}.

This paper is structured as follows. Section 2 provides a foundational introduction to dynamical systems theory, Kosambi-Cartan-Chern theory, and discusses the existing relation between these mathematical tools. In Section 3 provides an overview of Chern-Simons modified gravity and the slow rotation approximation to black holes within this framework. In Section 4, we present a dynamical system analysis of the conditions under which circular orbits remain stable in this modified gravity scenario, deepening our understanding of the dynamics of black holes within dCS. Section 5 is devoted to a discussion of our main results, concluding that for the system under consideration, the Lyapunov and KCC stability criteria are in complete agreement.
\section{Mathematical foundations}\label{sec2}

The purpose of this section is to provide the fundamental mathematical framework necessary to analyze the stability of the problem in question, thereby enabling us to conduct a deep analysis of the underlying physics. This chapter is divided into three sections; the first focuses on the linear theory of dynamical systems, second on the non-linear Kosambi-Cartan-Chern theory, and in the third part we comment on the differences between the two approaches.
\subsection{Elements of Lyapunov stability}

 In this section, only the key results that are directly applicable to our analysis are presented. For a more comprehensive understanding of the topic, additional references are suggested for further reading \cite{Perko,Verhulst,KuznetzovEOABT,Kl1,Kl2}. We are interested in differential equations expressed in the form 
\begin{equation}
    \dot{x}=f(x)\label{system},
\end{equation}where $x\in\mathbb{R}^n$, $E\in R^n$ and $f: E\to \mathbb{R}^n$. Critical points $x_c$ of equation \eqref{system} are those that satisfy $f(x_c)=0$.  A second order differential equation of the form 
\begin{equation} \ddot{y}+a\dot{y}+by=0,\label{eq2orden}
\end{equation}
where $a, b$ are constants, can be written in vector form using the transformation $x_1=y$ and $x_2=\dot{x}_1=\dot{y}$. Then the differential equation \eqref{eq2orden} is equivalent to the system of equations 
\begin{equation}
   \left.\begin{array}{lcl}
     \dot{x}_1=x_2
    ,
     \\\dot{x}_2=-bx_1-ax_2
   \end{array}\right\}\label{ODEsystem}
.\end{equation}
Equation \eqref{ODEsystem} is in the form of eq. \eqref{system} with $x=(x_1,x_2)^T$, i.e., we are in $\mathbb{R}^2$. If $x(t)=(x_1(t),x_2(t))$ is a solution of \eqref{ODEsystem} then $y=x_1(t)$ is a solution of \eqref{eq2orden}.  

 The critical points correspond to equilibrium solutions of the system, that is, $x(t)=x_c$ satisfies the equation for all time. The study of these points and solutions in their neighborhood is of particular interest. 
Through the application of linearization around the critical points, equation~\eqref{system} can be approximated as a linear equation
\begin{equation}
    \dot{x}=Ax \label{linealeq}.
\end{equation}
The next theorem establishes the existence and uniqueness of solutions for this kind of systems (see, e.g.~\cite{Perko}).
\begin{theorem}[Fundamental theorem of linear systems]\label{tthm1}
Consider an operator $A$  in $\mathbb{R}^n$, then the solution of the problem with initial conditions 
\begin{equation}\dot{x}=Ax;\hspace{0.5cm}x(0)=x_0\in\mathbb{R}^{n},\end{equation}\textit{is:}
\begin{equation}
  x(t) =   e^{tA}x_0,
\end{equation}
and there are no other solutions.
\end{theorem}
 The algebraic method of diagonalization can be used to reduce the coupled system \eqref{linealeq} to an uncoupled linear system. For systems in $\mathbb{R}^2$, it can be shown that under a suitable linear transformation of coordinates, the system \eqref{linealeq} is equivalent to $\dot{x}=Bx$, where $B$ has one of the following forms,
 \begin{equation}
    B=\begin{pmatrix}\omega & 0  \\
            0 & \mu \\ 
        \end{pmatrix} , \hspace{0.5cm} B=\begin{pmatrix}\omega & -1  \\
          0 & \omega \\ 
        \end{pmatrix}, \hspace{0.5cm} B=\begin{pmatrix}a& -b  \\
            b & a \\ 
        \end{pmatrix} .
 \end{equation}
Following the fundamental theorem for linear systems, it can be shown that solutions of the initial value problem $\dot{x}=Bx$ with $x(0)=0$ are given respectively by 
\begin{equation}
x(t)=\begin{pmatrix}e^{\omega t} & 0  \\
            0 & e^{\mu t} \\ 
        \end{pmatrix}x_0 , \hspace{0.25cm} x(t)=e^{\omega t}\begin{pmatrix}1 & t  \\
          0 & 1 \\ 
        \end{pmatrix}x_0, \hspace{0.25cm} x(t)=e^{at}\begin{pmatrix}\cos( bt) & -\sin (bt)  \\
            \sin (bt) & \cos (bt) \\ 
        \end{pmatrix} x_0.
\end{equation}
As a consequence of this, from the eigenvalues of $A$ we can  obtain important qualitative information about the behavior of solution curves near critical points in the phase space of Eq.~\eqref{linealeq}. In the two dimensional case, the eigenvalues of $A$ can be written in terms of $\det A=\delta$ and $\rm{Tr\,} A$ as 
\begin{equation}\lambda_{12}=\frac{1}{2}(\rm{Tr\,} A\pm\sqrt{(\rm{Tr\,} A)^2-4\delta}).\label{ec eigenvlores}\end{equation}
  
\begin{figure}[htbp]
    \centering
    \includegraphics[width=0.7\linewidth]{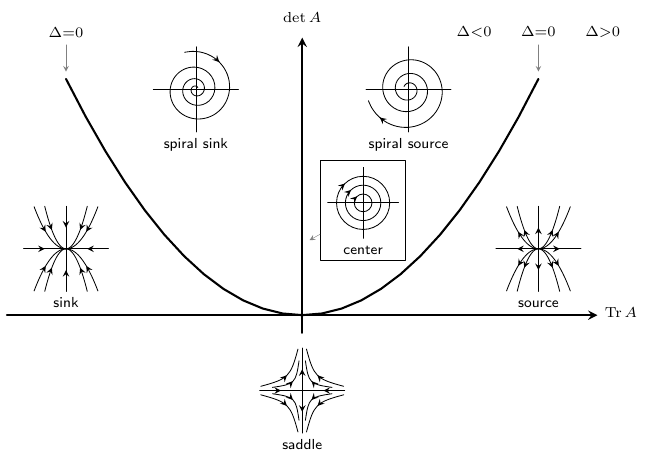}
    \caption{Classification of phase portraits for critical points in two dimensional systems depicted in terms of $detA$, $Tr A$ and $\Delta=Tr A-4\delta$. Based on figure 6, section 1.5 of  \cite{Perko}. }
    \label{figure1}
\end{figure}

  The phase portrait of the system \eqref{linealeq} is obtained from the phase portrait of $\dot{x}=Bx$ under a linear transformation of coordinates, in other words, the phase diagram of \eqref{linealeq} is topologically equivalent to those in Fig.~\ref{figure1}, and these results are summarized in Table~\ref{table1}. It should be noticed that this classification excludes cases where one or two eigenvalues are equal to 0; such critical points are called \textit{nonhyperbolic}, and require further analysis by determining the flow of the center manifold or using the center manifold theorem for higher-dimensional systems \footnote{The center manifold theorem is a fundamental result in the study of the stability of non-hyperbolic critical points, providing the necessary conditions to determine the behavior of the center manifold. See p. 196 of \cite{Verhulst}.}. For systems in $\mathbb{R}^3$, critical points have their three-dimensional counterparts of nodes, foci, centers, and so on. In the case where $\mathbb{R}^n$ with $n>3$, this classification is harder to visualize and the stability of systems is characterized in terms of attractive, repulsive, or periodic critical points and the existence of (un)stable manifolds~\cite{Perko,H/S}.

{ The Hartman-Grobman theorem is a fundamental result in the local qualitative theory of ordinary differential equations. It states that near a hyperbolic equilibrium point, the non-linear system \eqref{system} has the same qualitative behavior as the linearized system \eqref{linealeq} with $Df(x_0)=A$ (cf. \cite{Perko}).}

\begin{table}[htbp]
\caption{Classification of critical points for two dimensional systems. Based on main results of \cite{Perko,Verhulst}.}\label{table1}%
\begin{tabular}{@{}llll@{}}
\toprule
Eigenvalues&Classification of critical point&Stability\\
     \hline
 $\lambda_1,\lambda_2<0$ & Sink& Stable  \\
             $\lambda_1,\lambda_2>0$&Source & Unstable\\
        
        $\lambda_1<0<\lambda_2$&Saddle&Unstable\\
        
        $\lambda_{1,2}=a\pm ib, a<0$&Spiral sink&Stable\\
        
        $\lambda_{1,2}=a\pm ib, a>0$&Spiral source&Unstable\\ 
        
        $\lambda_{1,2}=\pm ib$&Center&Stable\\
        \botrule
\end{tabular}

\end{table}

\subsection{KCC Stability theory: Geometrization of arbitrary dynamical systems}
KCC theory is a mathematical framework for studying the stability of solutions of second-order differential equations. The central idea of KCC theory is that the evolution and behavior of the trajectories of the system of strongly nonlinear equations \eqref{ec.KCC 2 orden} can be analyzed using tools of the differential geometry of Finsler spaces~\cite{antonelli2014handbook, bao2012introduction, Harko_2016}; this is made possible by the analogy between paths in the Euler-Lagrange equations and geodesics in a Finsler geometry.

Finsler geometry is a generalization of Riemannian geometry, where the distance function, called the Finsler metric $ds=F(x, y)$, depends on both the position $x$ and the direction $y$ at each point. Unlike Riemannian geometry, which is isotropic (direction-independent) with a quadratic metric $ds=\sqrt{g_{ij}dx^{i}dx^{j}}$~\footnote{Riemannian geometry is a special case of Finsler geometry where the Finsler metric \( F(x,y) \) reduces to $F(x,y) = \sqrt{g_{ij}(x) y^i y^j}$. Here, the metric depends only on position $x$ and is quadratic in $y$, resulting in isotropic behavior.}, Finsler geometry facilitates the examination of anisotropic spaces, where distances can vary with direction; this flexibility makes Finsler geometry particularly useful for studying spaces with directional properties, such as those found in certain physical theories with preferred directions or complex movement costs. 

KCC theory~\cite{Antonelli2001,antonelli2014handbook,Antonellicelldivision2002} (we adopt Antonelli's notation) offers an alternative way to analyze the stability of systems with a more geometrical perspective, in which stability is described in terms of five invariants. Of these invariants, the second is particularly significant as it determines the stability of the system. These results are founded on the equivalence between the Euler-Lagrange equations and a system of second-order ordinary, strongly nonlinear differential equations as described by:
\begin{equation}
    \frac{d^{2}x^{i}}{dt^{2}}+2G^{i}(x^{j}, y^{j}, t)=0, \hspace{0.3cm} i=1,2,...,n,\label{ec.KCC 2 orden}
\end{equation}
where $G^i$ are the geodesic coefficients with $x^i=(x^1,x^2,...,x^n)$ and $y^i=(y^1,y^2,...,y^n)$  a set of dynamical variables that are defined in a real smooth $n$-dimensional manifold $\mathcal{M}$ and $t$ is the usual time coordinate. It is assumed that each function $G^i(x^j,y^j,t)$ is smooth in the neighborhood of some initial conditions $(x_0,y_0,t_0)$ defined in $\mathcal{TM}$~\footnote{The tangent bundle of $\mathcal{M}$ is denoted by $\mathcal{TM}$. Usually $\mathcal{M}$ is considered as $\mathbb{R}^n$, $\mathcal{M}=\mathbb{R}^n$, and therefore $\mathcal{TM}=\mathcal{T}\mathbb{R}^n=\mathbb{R}^n$. 
}. The geodesic coefficients $G^i$ are analogous to the geodesic spray coefficients in the Finsler geometry~\footnote{Spray is a special vector field on the tangent bundle $\mathcal{TM}$ that encodes the geodesics of the Finsler manifold.}. This geometrical approach allows for a deeper understanding of the stability properties of dynamical systems, particularly in contexts where anisotropic or directional dependencies play a critical role.

The next equations represent the central mathematical result of the KCC theory. The KCC-covariant derivative of an arbitrary vector field  $\chi$ is defined in terms of the second KCC invariant $P^i_j$ as
\begin{equation}
    \frac{D^{2}\chi^{i}}{dt^{2}}=P^{i}_j \chi^{j},\label{kcc-jacobi covariant}
\end{equation}
where
\begin{equation}
    P^{i}_j = -2\frac{\partial G^{i}}{\partial x^{j}} - 2G^{l}G^{i}_{jl} + y^{l}\frac{\partial N^{i}_j}{\partial x^{l}} + N^{i}_lN^{l}_j + \frac{\partial N^{i}_j}{\partial t},\label{kcc. curvature tensor}\end{equation}
the non-linear connection $N^i_j$ is defined as
\begin{equation}
    N^{i}_{j}=\frac{\partial G^{i}}{\partial y^{j}},
\end{equation} 
and the Berwald connection is
\begin{equation}
        G^{i}_{jl}\equiv \frac{\partial N^{i}_j}{\partial y^{l}}.
    \label{kcc.berwald connection}\end{equation}
The second KCC invariant $P^i_j$ is also called the deviation curvature tensor~\footnote{The deviation tensor measures the infinitesimal separation between nearby geodesics and is influenced by the directional curvature properties of the Finsler space.
}. It is the fundamental quantity in the KCC-theory and in the Jacobi stability method. If we assume that the system \eqref{ec.KCC 2 orden} corresponds to the geodesic motion of a physical system, then Eq.~\eqref{kcc-jacobi covariant} gives the Jacobi field equation.

The trace $P=P^i_i$ of the curvature deviation tensor is a scalar invariant and can be calculated from
\begin{equation}
    P=P^{i}_i=-2\frac{\partial G^{i}}{\partial x^{i}}-2G^{l}G^{i}_{il}+y^{l}\frac{\partial N^{i}_i}{\partial x^{l}}+N^{i}_lN^{l}_i + \frac{\partial N^{i}_i}{\partial t}.
\label{kcc.traza tensor curvatura}
\end{equation}
The analysis of the deviation tensor is essential for understanding the Jacobi stability or instability of geodesics, as it reveals how perturbations to geodesics behave over time in a direction-sensitive manner.

Other important invariants can also be constructed and are introduced according to the definitions 
\begin{equation}
    P^{i}_{jk}\equiv \frac{1}{3} \left(\frac{\partial P^{i}_j}{\partial y^{k}} - \frac{\partial P^{i}_k}{\partial y^{j}}\right),
    P^{i}_{jkl}\equiv \frac{\partial P^{i}_{jk}}{\partial y^{l}},
    D^{i}_{jkl}\equiv \frac{\partial G^{i}_{jk}}{\partial y^{l}}.
\end{equation} 
From a geometrical perspective $P^i_{jk}$ can be described as a torsion tensor, $P^i_{jkl}$ represents the equivalent of the Riemann-Christoffel curvature tensor and $D^i_{jkl}$ is the Douglas tensor.

The stability of trajectories in the vicinity of stationary solutions of a system of differential equations \eqref{ec.KCC 2 orden}, assuming that these trajectories represent smooth curves in the Euclidean space $\mathbb{R}^n$, can be determined by the following condition:
\begin{align}
    P^i_{j} & < 0 \quad \text{(stable)},\\
    P_{j}^{i} & > 0 \quad \text{(unstable)}.    
\end{align} 
If the first condition is satisfied, the trajectories are designated as Jacobi stable if and only if the real parts of the eigenvalues of the curvature deviation tensor $P^i_j$ are strictly negative. 
On the other hand, if the real parts of the eigenvalues of the deviation curvature tensor $P^i_j$ are strictly positive, the trajectories of the dynamical system are designated as unstable.
This definition is an alternative to the standard linear analysis to investigate the stability of second-order differential equations.
\subsection{Comments about stability}
In this section, we briefly compare linear (Lyapunov) stability with KCC (Jacobi) stability, with the goal of establishing a more robust framework that fortifies the theoretical foundations of dynamical systems. For a more detailed discussion~\cite{boemer2010, abolghasem, Paul, Strogratz}.
As mentioned above, various types of stability can be considered when talking about dynamical systems. Lyapunov theory covers stability near critical points $x_c$ of \eqref{system} and is one of the most important ones. If all trajectories that begin near $x_c$ approach this point as $t\to\infty$, i.e., $\lim_{t\to \infty}x(t)=x_c\hspace{0.1cm}\forall \hspace{0.1cm}\|x(0)-x_c\|<\delta$, then we say that every trajectory beginning within a distance $\delta$ from $x_c$ eventually converges to it.

A different notion of stability takes into account the behavior of trajectories for all $t$, not only when $t\to\infty$. The point $x_c$ is \textit{Lyapunov stable} if all trajectories that begin near $x_c$ remain near $x_c$ at all times. There can be critical points that are Lyapunov stable but not attractors. This situation occurs frequently and the point $x_0$ is called \textit{ neutrally stable}. Trajectories near $x_c$ are neither attracted nor repealed from a neutrally stable point. Finally, a point $x_c$ is \textit{asymptotically stable} if it is an \textit{attractor} and \textit{Lyapunov stable}. In contrast, Jacobi stability, as studied in KCC theory, takes a global and geometric approach, examining trajectories as geodesics in a Finsler space, evaluating how small perturbations influence the system's entire path via the deviation tensor. Jacobi stability examines the robustness of the system to parameter changes and considers non-linear effects across a broader range of the system's behavior.

Lyapunov stability analysis and KCC theory both provide valuable insights into the stability of dynamical systems under perturbations, but they approach stability from distinct perspectives. Although different in scope -- Lyapunov stability being local and algebraic and Jacobi stability being global and geometric -- these methods complement each other. Lyapunov stability identifies local conditions for stability around equilibrium points, while the Jacobi stability provides a broader view of trajectory behavior and long-term robustness. Together, they offer a comprehensive framework for understanding system stability, combining local precision with global insights.

In conclusion, Lyapunov stability and Jacobi stability serve as complementary methodologies for assessing the stability of dynamical systems. By integrating both local and global perspectives, these approaches provide a more profound understanding of the system's behavior, which is crucial for precise stability analysis in both physical and theoretical contexts.

\section{Formulation of Chern-Simons modified gravity}\label{section3}
Chern-Simons modified gravity (CS) is a four-dimensional extension of General Relativity (GR) that covers a variety of theories, each characterized by specific couplings $\alpha$ and $\beta$. Two significant formulations are identified: the non-dynamical framework, where $\beta = 0$; and the dynamical framework, where both $\alpha$ and $\beta$ are arbitrary but nonzero. In this analysis, we consider the dynamical framework and utilize the notation of~\cite{Yunes_2009,Alexander_2009}. 
The action defining this theory is 
\begin{equation}\label{actioncs}
S_\text{total}=S_\text{EH}+S_\text{CS}+S_{\vartheta}+S_\text{M}.
\end{equation}
The first term corresponds to the Einstein-Hilbert action,
\begin{equation}
S_\text{EH}= \kappa \int d^{4}x\sqrt{- g }R,
\end{equation}
where $\kappa=\left ( 16 \pi\right)^{-1}$ is the dimensionless gravitational coupling constant ($c$ and $G$ are normalized to 1,) $R$ is the Ricci scalar, and $g$ represents the determinant of the metric tensor $g_{\mu\nu}$.

The expression $S_\text{CS}$ for the CS action is\footnote{Couplings mediated by an arbitrary function of $\vartheta$ could also be considered. However, the approximate analytic spinning black hole solution that is of interest for us is known only for the linear coupling and vanishing potential $V(\vartheta)=0$ in Eq.~\eqref{eq:actsf}. Numerical, non-perturbative solutions for the same model have been explored in~\cite{Delsate:2018ome}, finding good agreement up to $\zeta$ of order one. More general couplings have been studied in the context of spontaneous scalarization~\cite{Richards:2025ows,Doneva:2021dcc} and primordial gravitational waves~\cite{Feng:2023veu,Bartolo:2018elp}. In the latter case, no significant differences are found between linear and more general couplings. }
\begin{equation}
S_\text{CS} = \frac{\alpha}{4}\int d^{4}x \sqrt{-g}\vartheta {}^{\star}\!R R,
\end{equation}
where $\alpha$ represents the CS coupling constant measured in units of length squared, $\vartheta$ is a dimensionless scalar field referred to as the Chern-Simons coupling field and ${}^{\star}\!R$ denotes the dual Riemann tensor that constructs the Pontryagin density 
\begin{equation}\label{pontry}
    {}^{\star}R R = {}^{\star}R^{\mu}{}_{\nu}{}^{\rho \sigma} R^{\nu}{}_{\mu}{}_{\rho \sigma} = \frac{1}{2}\epsilon^{ \rho \sigma \delta\tau } R^{\mu}{}_{\nu}{}_{\delta\tau}R^{\nu}{}_{\mu}{}_{\rho \sigma}.
\end{equation}

with $\epsilon^{ \rho \sigma \delta \tau }$ being the 4-dimensional Levi-Civita tensor.

The action for the scalar field is expressed as
\begin{equation}\label{eq:actsf}
S_{\vartheta}=-\frac{\beta}{2}\int d^{4}x\sqrt{-g}\left[g^{\mu\nu}\nabla_{\mu}\vartheta \nabla_{\nu}\vartheta +2V(\vartheta) \right],
\end{equation}
where $\nabla_{\mu}$ represents the covariant derivative, $\beta$ denotes a dimensionless coupling constant, and $V(\vartheta)$ is a potential for the scalar field, which is set to zero in the rest of this work. Finally, the matter action $S_M$ is also set to zero.

When the scalar field $\vartheta$ is constant, CS modified gravity becomes equivalent to GR. This happens because the Pontryagin term~\eqref{pontry} can be written as a divergence  of the Chern-Simons topological current,
\begin{equation}
\nabla_{\mu}K^{\mu}=\frac{1}{2}{}^{\star}R R,
\end{equation}
 where $ K^{\mu}$ is given by
\begin{equation}
    K^{\mu}=\epsilon^{ \mu \rho \sigma \tau }\Gamma^{\upsilon}_{\rho\psi}\left ( \partial_{\sigma}\Gamma^{\psi}_{\tau \upsilon}+\frac{2}{3}\Gamma^{\psi}_{\sigma\xi}\Gamma^{\xi}_{\tau \upsilon} \right ),
\end{equation}
with $\Gamma^{\psi}_{\sigma\xi}$ representing the Christoffel symbols.

The field equations are derived by varying the action with respect to the metric and the CS coupling field. The variation with respect to the metric results in
\begin{equation}
G_{\mu\nu}+\frac{\alpha}{\kappa}C_{\mu\nu}=\frac{1}{2\kappa}T_{\mu\nu},
\end{equation}
where $G_{\mu\nu}$ is the Einstein tensor and $C_{\mu\nu}$ is the trace-free C-tensor~\footnote{Brackets around the indices indicate symmetrization, for instance, $B_{(\mu\nu)}:=(B_{\mu\nu} + B_{\nu\mu})/2$.} 
\begin{equation}
    C^{\mu\nu} = \left( \nabla_{\sigma}\vartheta \right )\epsilon^{\sigma \delta \alpha ( \mu}
\nabla_{\alpha}R^{\nu)}{}_{\delta}+\left (\nabla_{\sigma}\nabla_{\delta}\vartheta \right ) {}^{\star}\!R^{\delta (\mu \nu )\sigma}.
\end{equation}
Variation with respect to the CS coupling field results in the Klein-Gordon equation for a massless scalar field with a source term,
\begin{equation}
    \beta\Box\vartheta=-\frac{\alpha}{4}\,{}^{\star}\!R R,
\end{equation}
where $\Box:=g^{\mu\nu}\nabla_{\mu}\nabla_{\nu}$ is the d’Alembert operator. 

\subsection{Black holes in dCS modified gravity: slow rotation approximation}\label{sec:hjcs} 
Adding a CS term to the Einstein-Hilbert action modifies the rotating solution of General Relativity~\cite{Jackiw_2003,Yunes_2009,Konno_2007,Konno_2009}. The metric for the solution in the modified theory under the slow rotation approximation is given by
 \begin{equation}\label{a5}
 ds^{2}= ds^{2}_\text{Kerr}+ \frac{5}{4} \frac{\alpha^{2}}{\beta \kappa} \frac{a}{r^{4}} \left(1+\frac{12}{7}\frac{m}{r}+ \frac{27}{10}\frac{m^{2}}{r^{2}} \right)\sin^{2}{\theta} d\phi dt,
 \end{equation}
 where in Boyer-Lindquist coordinates the geometry of the Kerr metric is~\cite{visser2008kerrspacetimebriefintroduction,Chandrasekhar1985kt}
 \begin{align}
    ds^{2}_\text{Kerr} = &-\left(1 - \frac{2mr}{\rho^2} \right) dt^{2} 
    - \left(\frac{4mr\sin^{2}{\theta}}{\rho^2}\right) dtd\phi 
    + \left(\frac{\rho^2}{\Delta}\right) dr^2 
    + \rho^2 d\theta^2 \nonumber \\ 
    &+ \sin^{2}{\theta} \left(r^{2} + a^{2} + \frac{2mar\sin^{2}{\theta}}{\rho^2}\right) d\phi^{2},
\end{align}
where
\begin{align}
    \rho^2 &= r^{2} + a^{2} \cos^2{\theta}, \\
    \Delta &= r^{2} + a^{2} - 2mr.
\end{align}
 The scalar field compatible with this solution is
\begin{equation}
     \vartheta = \frac{5\alpha a \cos{\theta}}{8\beta m r^{2}}\left( 1+\frac{2m}{r}+ \frac{18m^{2}}{5r^{2}} \right).
 \end{equation}
 
Notice that the modified metric remains asymptotically flat at infinity. In equation~\eqref{a5}, $ds^{2}_\text{Kerr}$ represents the Kerr metric in the slow rotation approximation, where $m$ denotes the black hole's geometrized mass and $a$  its specific angular momentum. The constants $\alpha$ and $\beta$ are the coupling constants that appear in the action.

 In the following analysis, we simplify the expression in Eq.~(\ref{a5}) by neglecting the second and third terms enclosed by parentheses. This approximation is justified because, for small rotation and small CS coupling parameters, we expect the smallest circular matter orbit to be near $r\sim 6m$, where the terms that we are neglecting are already smaller than the first term in the parentheses.  
 Hence, we use the metric
 \begin{equation}\label{approx}
 ds^{2}= ds^{2}_\text{Kerr}+ \frac{5}{4} \frac{\alpha^{2}}{\beta \kappa} \frac{a}{r^{4}} \sin^{2}{\theta} d\phi dt,
 \end{equation}
 and perform our calculations primarily at leading order.

 For the purpose of testing the theory, it is beneficial to define the parameter $\xi$ which is measured in units of $[\text{length}]^{4}$, and its dimensionless version $\zeta$, 
\begin{align}
    \xi &:= \frac{\alpha^{2}}{\beta \kappa}, \ \ \ 
    \zeta:= \frac{\xi}{m^{4}}.
\end{align}
The solution for a rotating black hole in dCS gravity simplifies to the Kerr metric when $\alpha=0$, or equivalently when $\xi=0$ or $\zeta=0$. In this work, the values of $\xi$ are restricted by the conditions delineated in \cite{rodríguez2024shadowsblackholesdynamical}, where compatibility with the black hole shadow reported by the EHT~\cite{2019m87,Event_Horizon_Telescope_Collaboration_2022} is analyzed.

\section{Stability Analysis for timelike geodesics in dCS}\label{section4}

In this section, we perform a stability analysis of the equations using the mathematical frameworks reviewed in the previous sections. To precisely compute the pertinent physical quantities, starting from the one-dimensional geodesic equation, we initially determine the radial dependence of the angular velocity $\Omega$, the specific energy $E$, and the specific angular momentum $L$ of particles in circular orbits~\cite{Harko_2009, Harko_2010}, and we carry out an examination of the equatorial plane $\theta = \frac{\pi}{2}$. The latter manifests as follows:

\begin{equation}
    g_{rr}\left( \frac{dr}{d\tau}\right)^{2} = -1+\frac{E^{2}g_{\phi\phi}+2ELg_{t\phi}+L^{2}g_{tt}}{g^{2}_{t\phi}-g_{tt}g_{\phi\phi}}.
\end{equation}
From this equation, we can identify the effective potential for timelike geodesics in the context of the Chern-Simons slow rotation approximation~\eqref{approx} as
\begin{equation}
    V_\text{Eff}(r)=\frac{g_{t\phi}^2 - E^2 g_{\phi\phi} - 2 E g_{t\phi} L - g_{tt} (g_{\phi\phi} + L^2)}{g_{rr} (g_{t\phi}^2 - g_{tt} g_{\phi\phi})},
\end{equation}
such that 
\begin{equation}
    \ddot{r}=-\frac{1}{2}V'_\text{Eff}(r).\label{Ec2orden r potencial}
\end{equation}
In this specific instance, we factored out the constant ${1}/{2}$ from the effective potential since it does not influence the Lyapunov analysis nor the Jacobi analysis. However, it is recommended that future studies take this factor into consideration.
The effective potential is decomposed into a Kerr-related component and a Chern-Simons correction term, 

\begin{align}
    V_\text{Eff}(r) &= V(r)_\text{K} + \xi V(r)_\text{CS}, \\
    V(r)_\text{K} &= 1 - \frac{2 (-a E + L)^2m + a^2 (-1 + E^2) r - L^2 r + 2 m r^2 + E^2 r^3}{r^3}, \\
    V(r)_\text{CS} &= -\frac{5 a \big(2 a E m + L (-2 + r)\big) \big(-2 a L m + E r^3 + 
    a^2 E (\mathcal{F}_{-2})\big)}{4 r^8 \big(a^2 + r \mathcal{F}_{2}\big)}.
\end{align}\label{potenciald}
with the functions $\mathcal{F}$ and $\mathcal{F}_{-2}$ defined as follows, 
\begin{align}
     \mathcal{F}_{n} & =r-nm. \\
     \mathcal{F}^{*}_{n} & =(n-5)r-nm. \\
    \mathcal{F}^{**}_{n} & =3nm^{2}-3(n+1)mr-nr^{2}.
\end{align}
 where we have also defined the functions $\mathcal{F}^{*}_{n}$ and $\mathcal{F}^{**}_{n}$ for future use.
 
\begin{figure}[htbp]
\centering

 \includegraphics[width=0.3\textwidth]{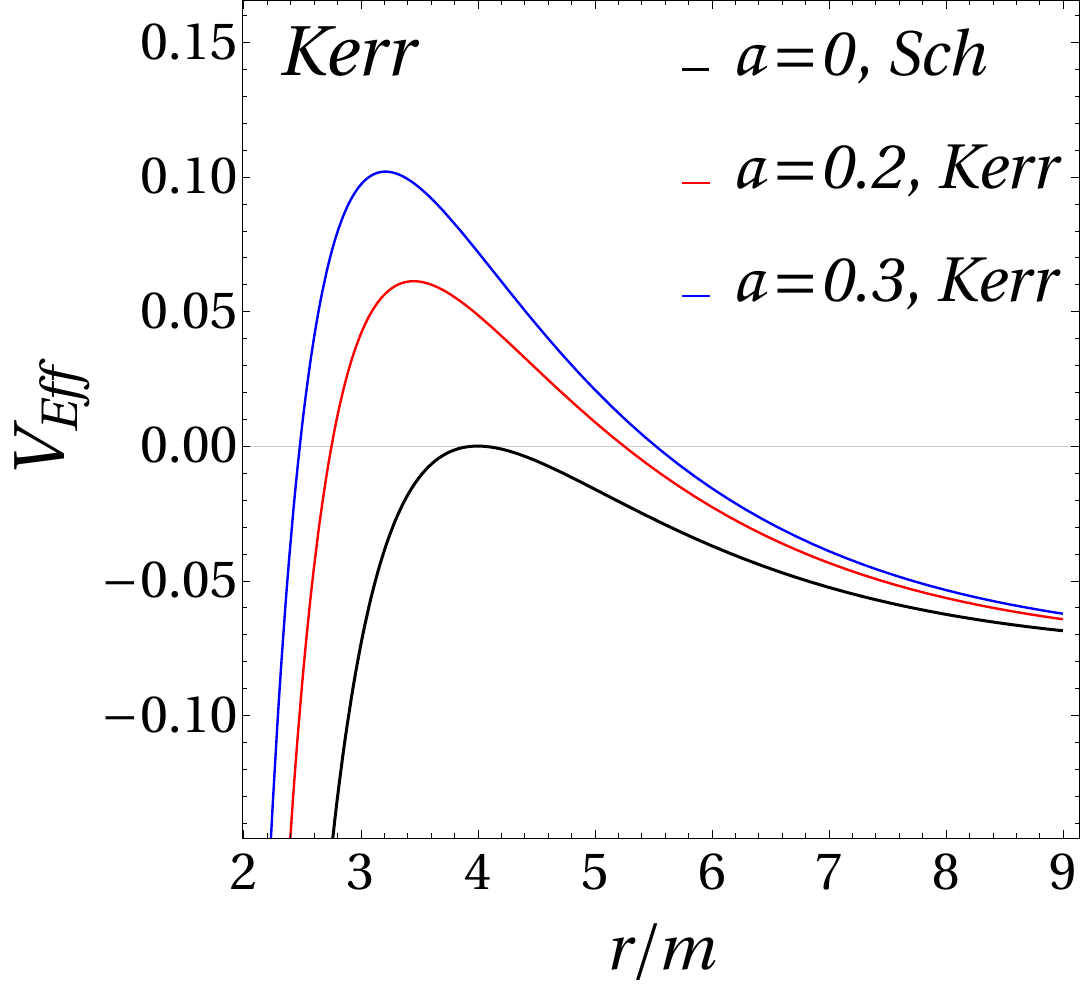}
\includegraphics[width=0.3\textwidth]{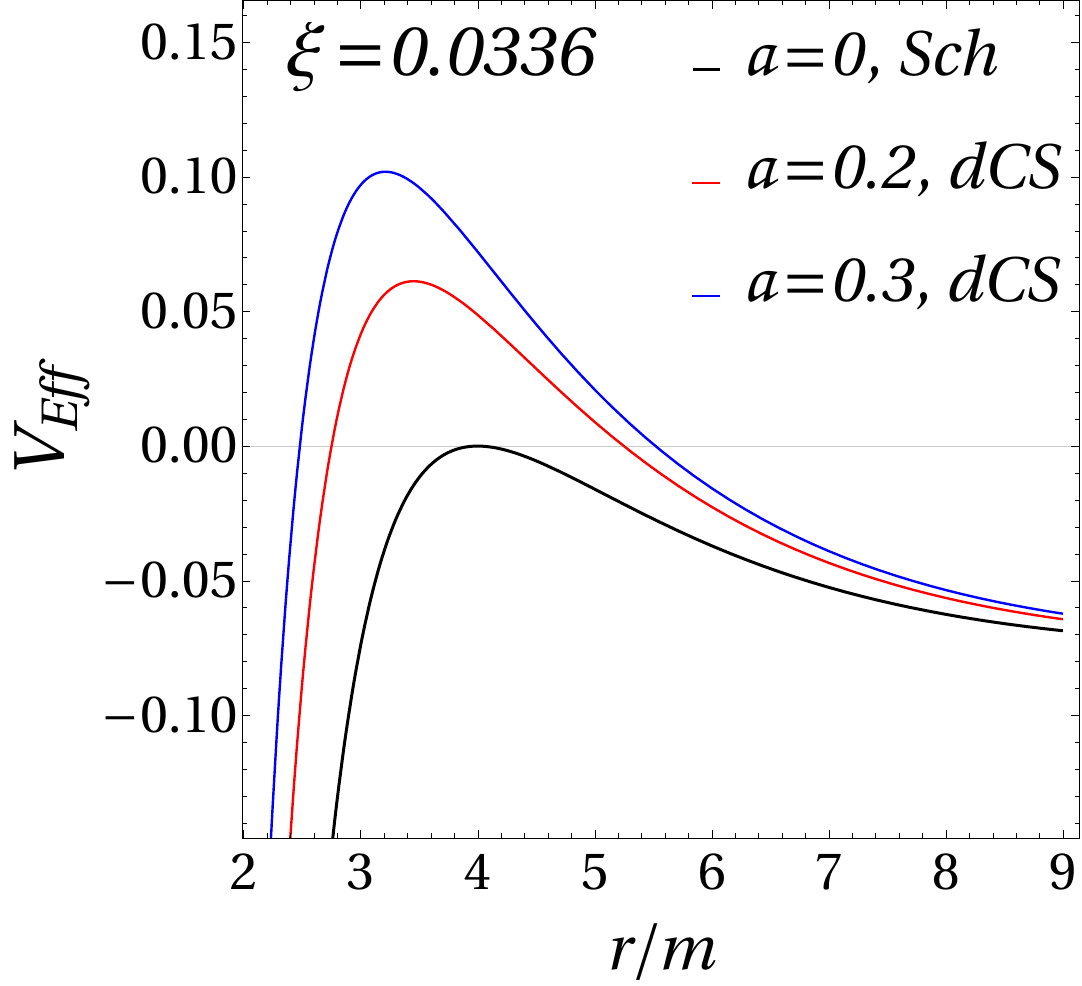}
\includegraphics[width=0.3\textwidth]{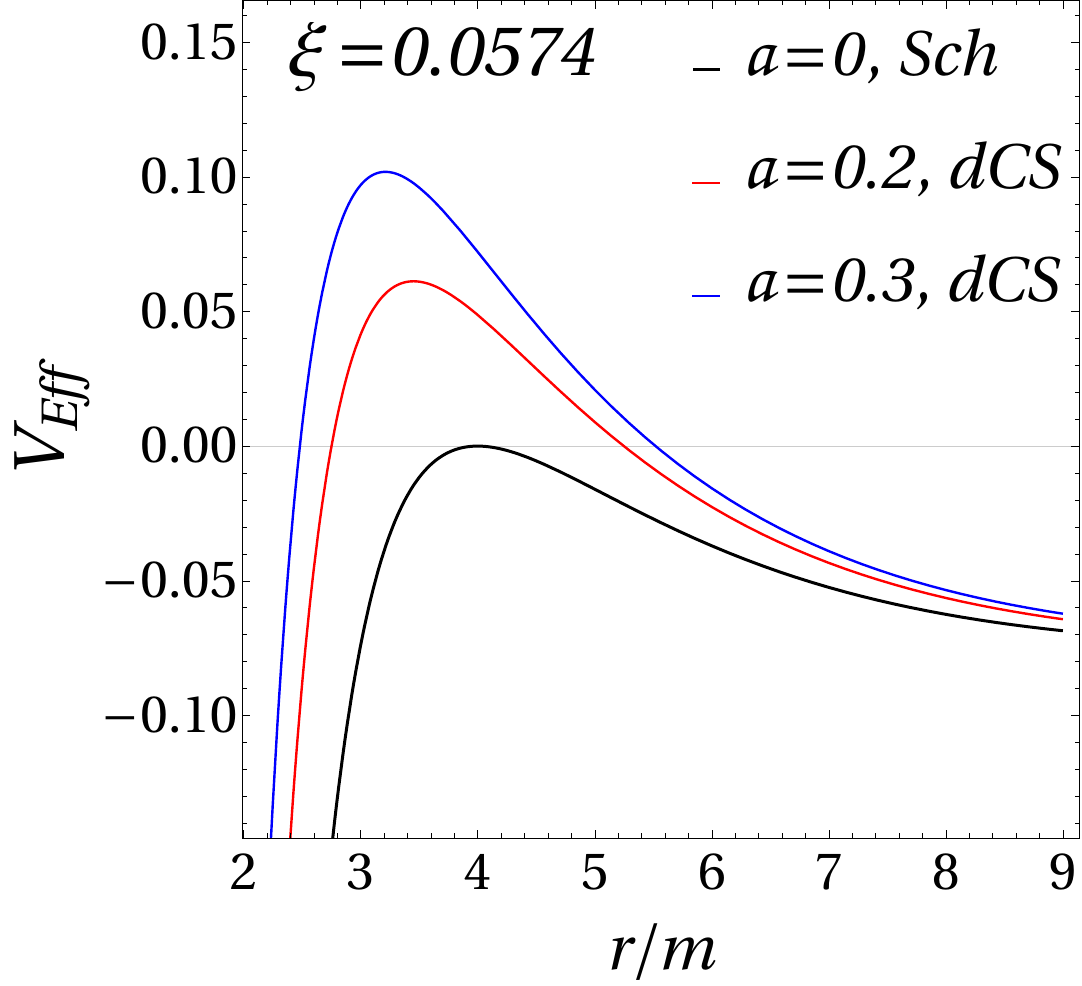}

\caption{The effective potential for GR and dCS spacetimes concerning timelike geodesics characterized by various spin parameters $a$ and $\xi$.}
\label{fig:CSdVeff}
\end{figure}

\subsection{Linear stability}
With the purpose of analyzing the stability of equation \eqref{Ec2orden r potencial} a system of equations can be constructed in $\mathbb{R}^2$ using the coordinate transformation $\dot{r}=p$ and $\dot{p}=\ddot{r}$,
\begin{equation}
   \begin{array}{lcl}
     \dot{r}=p(r)
    ,
     \\\dot{p}=-V'_\text{Eff}(r),
   \end{array}\label{DS chs}
\end{equation}
where
\begin{align}
    V'_\text{Eff}(r) &= V'(r)_\text{K} + \xi V'(r)_\text{CS}, \\
    V'(r)_\text{K} &= \frac{6 (-a E + L)^2 m + 2 r \big(a^2 (-1 + E^2) - L^2\big) + 2 m r^2}{r^4}, \\
    V'(r)_\text{CS} &= \frac{15 a E L r^3 + 5 a^2 M \left(-9 L^2 + \frac{E^2 \mathcal{F}^{*}_{12} r^3}{\mathcal{F}_{2}^2}\right)}{2 r^{10}}.
\end{align}
Eqs. \eqref{DS chs} may be reformulated in vector notation with $x=(r,p)^T$, thereby constituting a nonlinear differential equation of the form $\dot{x}=f(x)$, where 
\begin{equation}
f(x,t)=\begin{pmatrix} p \\
-V'_\text{Eff}(r)
\end{pmatrix}.\label{VecFun}
\end{equation}
The critical points $x_c$ of the system are determined by condition $f(r,p)=0$, i.e.,  $p=0$ and $V'_\text{Eff}(r)=V'(r)_\text{K}+\xi V'(r)_\text{CS}=0$.  

Solving for $r$ the critical radius values $r_{*}$ are found as $r_{*}=r_\text{K*}+\xi r_\text{CS*}$, where~\footnote{The critical points by definition are real, so only the real radii are taken.}

\begin{align}
   r_\text{K*}&= \frac{-a^2 (1 - E^2) + L^2 \mp \sqrt{(a^2 - a^2 E^2 + L^2)^2 - 12 (-a E + L)^2 m^2}}{2 m}, \\
r_\text{CS*}&=\frac{5 a \left(3 E L \, r_\text{K*}^3 (\mathcal{F}_\text{2K*})^2 + a m \left(-9 L^2 (\mathcal{F}_\text{2K*})^2 + E^2 r_\text{K*}^3 (\mathcal{F}^{*}_\text{12K*})\right)\right)}{4 \, r_\text{K*}^5 (\mathcal{F}_\text{2K*})^2 \left(12 (-a E + L)^2 m + 3 a^2 (1 - E^2) r_\text{K*} - 3 L^2 r_\text{K*} + 2 m r_\text{K*}^2\right)}.
\end{align}
The subscript $K^*$ in the functions $\mathcal{F}$ indicates that these are evaluated at $r_\text{K*}$. 

The system exhibits critical points $x_{c_i} = (0, r_{*})$, around which stability analyzes are performed. For these analyzes, and by the Hartman-Grobman theorem, the derivative of the function $f(r, p)$, specifically the Jacobian matrix, is required, 
\begin{align}
    Df(x,t) = 
    \begin{pmatrix} 
        \frac{\partial f_1}{\partial r} & \frac{\partial f_1}{\partial p} \\
        \frac{\partial f_2}{\partial r} & \frac{\partial f_2}{\partial p}
    \end{pmatrix} = 
    \begin{pmatrix}
        0 & 1 \\
        -V''(r) & 0
    \end{pmatrix}.
\end{align}
The eigenvalues of this matrix, evaluated at $r_*$, indicate the stability of each point, respectively
\begin{equation}
    \lambda_{12}=\pm\sqrt{-V''_\text{Eff}(r_{*})}\label{Eigenvalues}.
\end{equation}
For the cases where $a=0$, $0.2$ and $0.3$ and $\xi=0$, $0.0336$ and $0.0574$, tables~\ref{tab2},~\ref{tab3} and~\ref{tab4} show the behavior of $r_{*}$ and $V''_{\text{Eff}}$  for fixed values of parameters $m=1$, $E=1$, $L=4$. Table~\ref{tab5} provides the range of stable time-like circular orbits. These radii are used to determine the innermost stable circular orbit (ISCO), which corresponds to the inner boundary of the accretion disk. The determination of the marginally stable orbit surrounding the central object can be obtained by the additional condition $V''_{\text{Eff}}(r) = 0$, which yields the subsequent significant relation~\cite{Harko_2009}
\begin{equation}
    E^{2}g''_{\phi\phi}+2E L g''_{t\phi}+L^{2}g''_{tt}-\left ( g^{2}_{t\phi}-g''_{tt}g''_{\phi\phi} \right )''=0,
\end{equation}
where the derivatives are with respect to $r$. By solving this equation for $r$, the radii of the marginally stable orbits are determined. For dCS gravity, the location of the ISCO is
\begin{equation}
   r_\text{ISCO}=6 m \mp 4 \sqrt{\frac{2}{3}} a - \frac{7 a^2}{18 m} \pm \frac{25 a \, \xi}{432 \sqrt{6} \, m^4},
\end{equation}
where the upper signs correspond to co-rotating geodesics, while the lower signs are for counter-rotating ones.

\begin{table}[htbp]
\caption{Values of $V''(r_{*})$ evaluated at critical radius $r_{*}$ for $a=0$ and different values of coupling parameter $\xi$.}\label{tab2}%
\begin{tabular}{@{}lllll@{}}
\toprule
 $\xi$  & $r_1$ & $r_2$ &$V''(r_1)$ & $V''(r_2)$\\ 
\midrule
0 & 4 &12 &-0.0625&0.00077\\
0.0336&4& 12 &-0.0625&0.00077\\ 
0.0574 &4 &12 & -0.0625& 0.00077\\
\botrule
\end{tabular}
\end{table}
\begin{table}[htbp]
\caption{Values of $V''(r_{*})$ evaluated at critical radius $r_{*}$ for $a=0.2$ and different values of coupling parameter $\xi$.}\label{tab3}
\begin{tabular}{@{}lllll@{}}
\toprule
 $\xi$  & $r_1$ & $r_2$ &$V''(r_1)$ & $V''(r_2)$\\ 
\midrule
0 & 3.45247&12.54752 &-0.12803&0.00073\\ 
0.0336&3.45275& 12.54752&-0.12801&0.00073\\ 
0.0574&3.45293 &12.54752 &-0.12800& 0.00073\\ 
\botrule
\end{tabular}
\end{table}
\begin{table}[htbp]
\caption{Values of $V''(r_{*})$ evaluated at critical radius $r_{*}$ for $a=0.3$ and different values of coupling parameter $\xi$.}\label{tab4}
\begin{tabular}{@{}lllll@{}}
\toprule
 $\xi$  & $r_1$ & $r_2$ &$V''(r_1)$ & $V''(r_2)$\\ 
\midrule
0 & 3.21147 &12.78852 &-0.18007&0.00072\\
0.0336&3.21194 & 12.78852 &-0.18003&0.00072\\ 
0.0574 &3.21228 &12.78851 & -0.17999& 0.00072\\
\botrule
\end{tabular}

\end{table}

\begin{figure}[ht]
     \centering
 
     \includegraphics[width=0.31\textwidth]{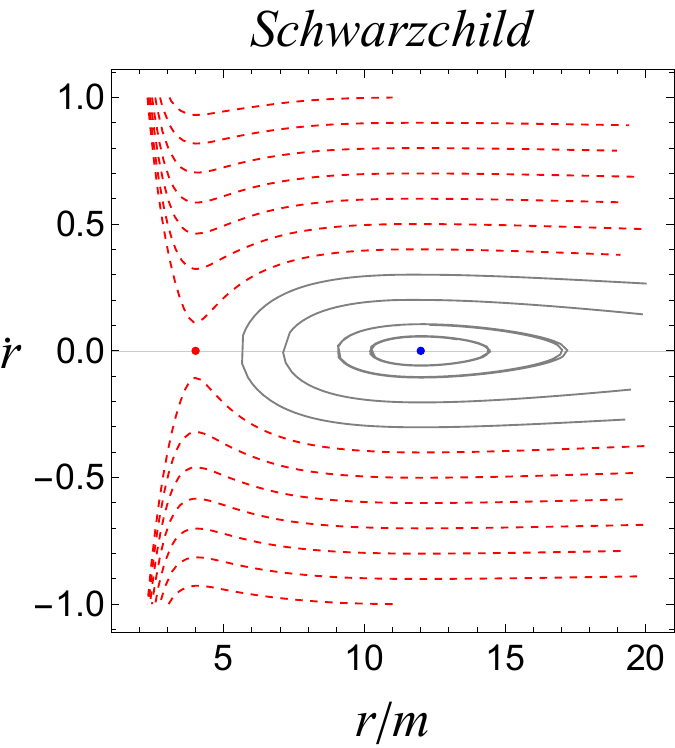}
   
     \caption{Phase diagram in the   $r-\dot{r}$  plane for timelike geodesics in Schwarzschild Spacetime, shown for spin parameter $a=0$, and coupling parameter value $\xi=0$; The red highlighted point $x_1=(0,r_1)$ is a saddle point, nearby solutions (dashed) in the neighborhood are unstable; the blue highlighted critical point $x_2=(0,r_2)$ is a stable center, and solutions near (solid) are stable closed orbits.}
     \label{fig3}
 \end{figure}

\begin{figure}[htbp]
\centering
 
\includegraphics[width=0.31\textwidth]{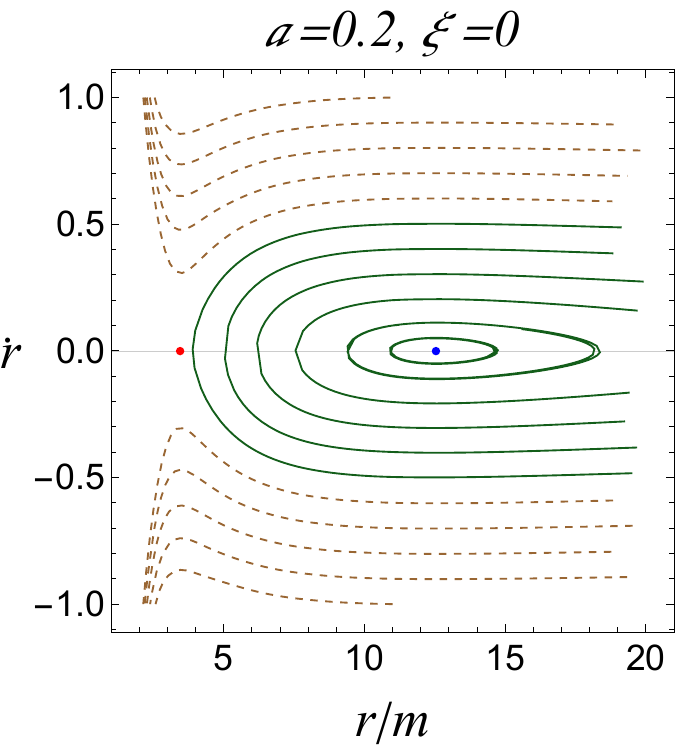} 
\includegraphics[width=0.31\textwidth]{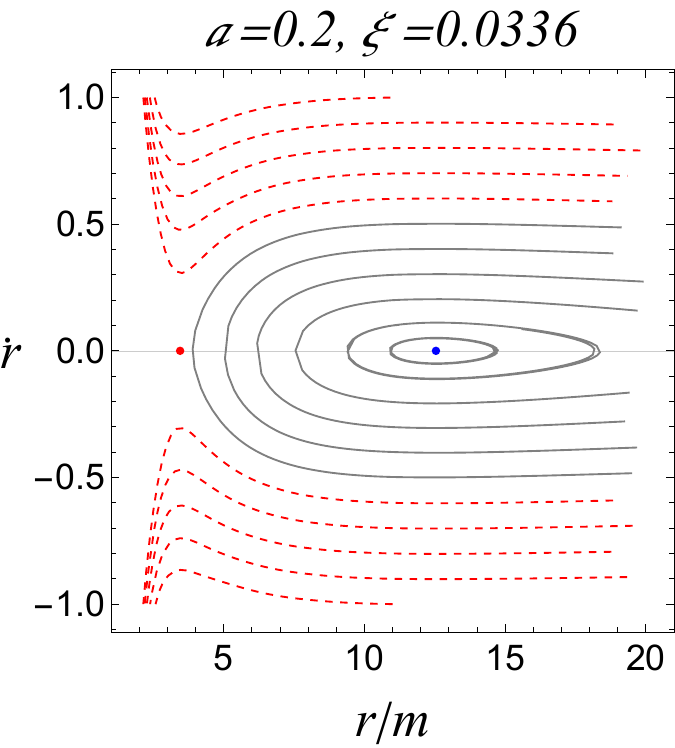}
\includegraphics[width=0.31\textwidth]{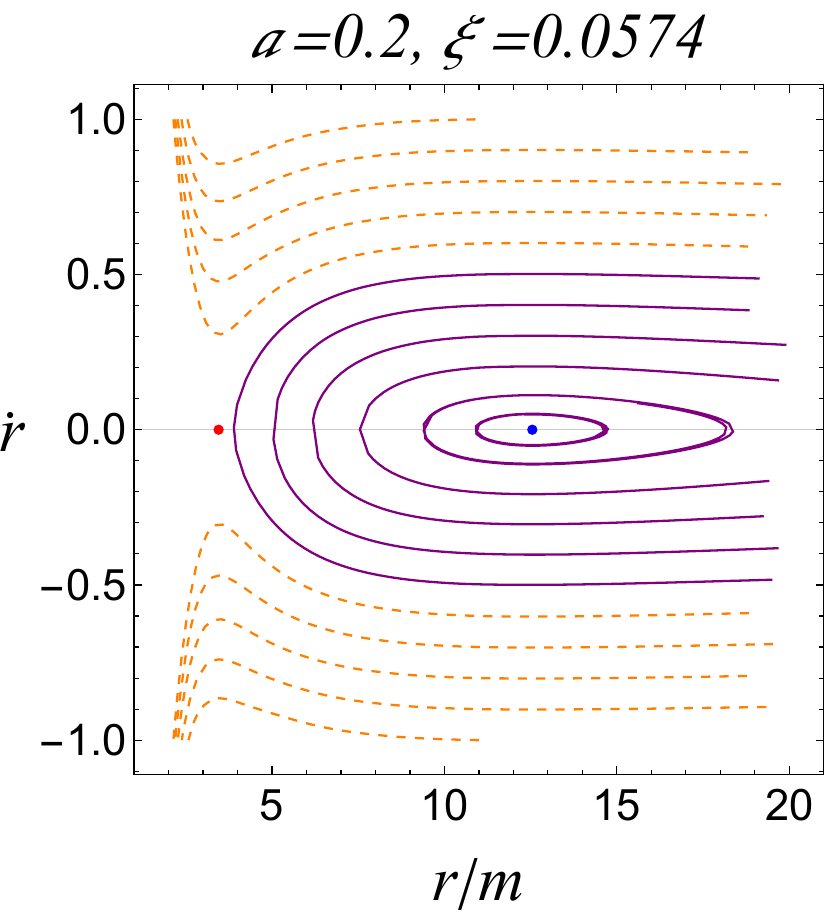}

  \caption{Phase diagrams in the   $r-\dot{r}$  plane for timelike geodesics in dCS Spacetime, shown for spin parameters from left to right $a=0, 0.2, 0.3$, and $\xi=0.0336$; The red highlighted point $x_1=(0,r_1)$ is a saddle point, nearby solutions (dashed red) in the neighborhood are unstable; the blue highlighted critical point $x_2=(0,r_2)$ is a stable center, and solutions (solid gray) near are stable closed orbits. It is observed that the critical points exhibit a slight displacement along the $r$ axis as variations occur in the spin parameter $a$.}
\label{fig4}
\end{figure}

The eigenvalues associated to the critical points are given by Eq.~\eqref{Eigenvalues}, therefore, from Table~\ref{tab2} we observe that there are two critical points. We observe that for $a=0$ and $\xi=0$, $\xi=0.0336$ and $\xi=0.0574$, at the critical point $x_1=(0,r_1)$ the eigenvalues are real and opposite in sign, thus $x_1$ is an unstable saddle point. At the critical point $x_2=(0,r_2)$, the eigenvalues are purely imaginary, therefore $x_2$ is a stable center. The phase portraits are sketched in Fig.~\ref{fig3}. Phase portraits exhibit similar behavior for $a=0.2$ and $a=0.3$, as can be observed in Fig.~\ref{fig4},  derived from the eigenvalues obtained from tables ~\ref{tab3} and \ref{tab4}.  Stability depends entirely on the parameters' values $a, E, L, m,$ and $\xi$.

\begin{figure}[htbp]
\centering
 
\includegraphics[width=0.31\textwidth]{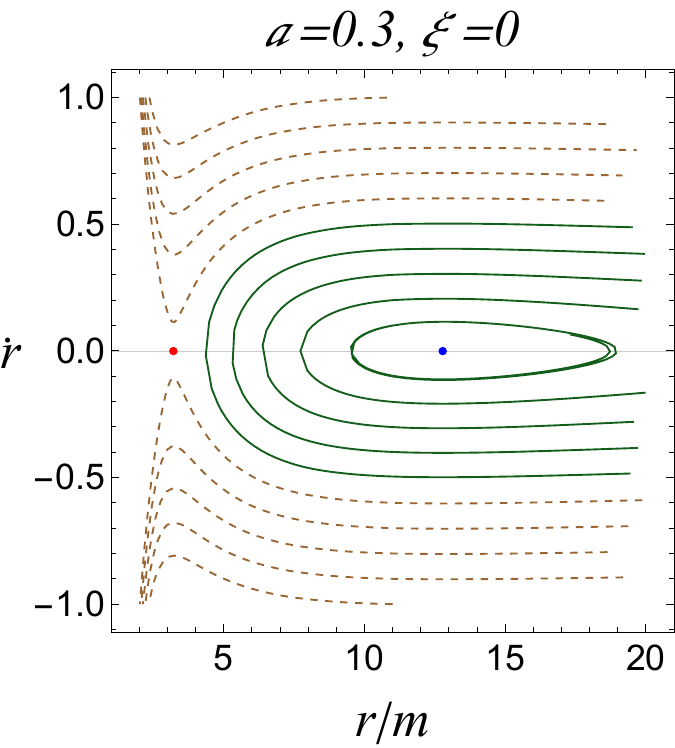}
\includegraphics[width=0.31\textwidth]{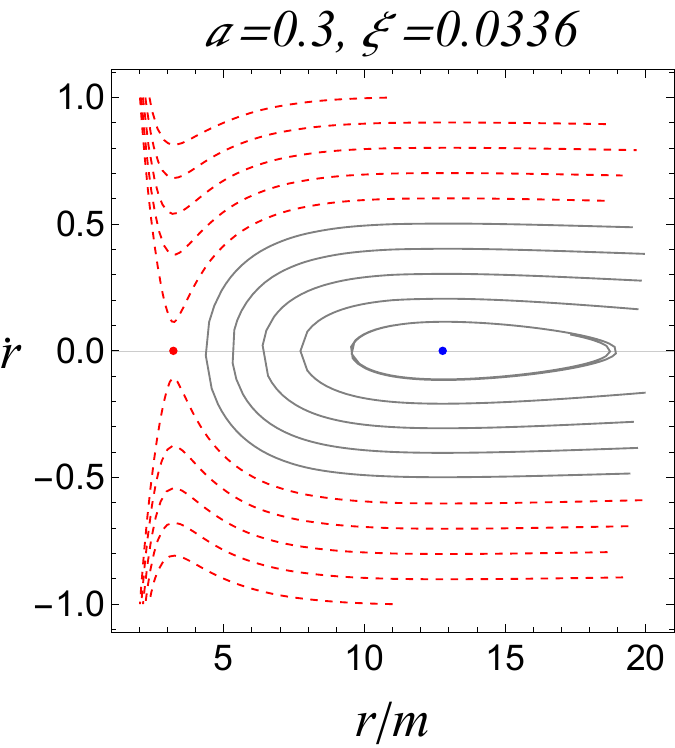}
\includegraphics[width=0.31\textwidth]{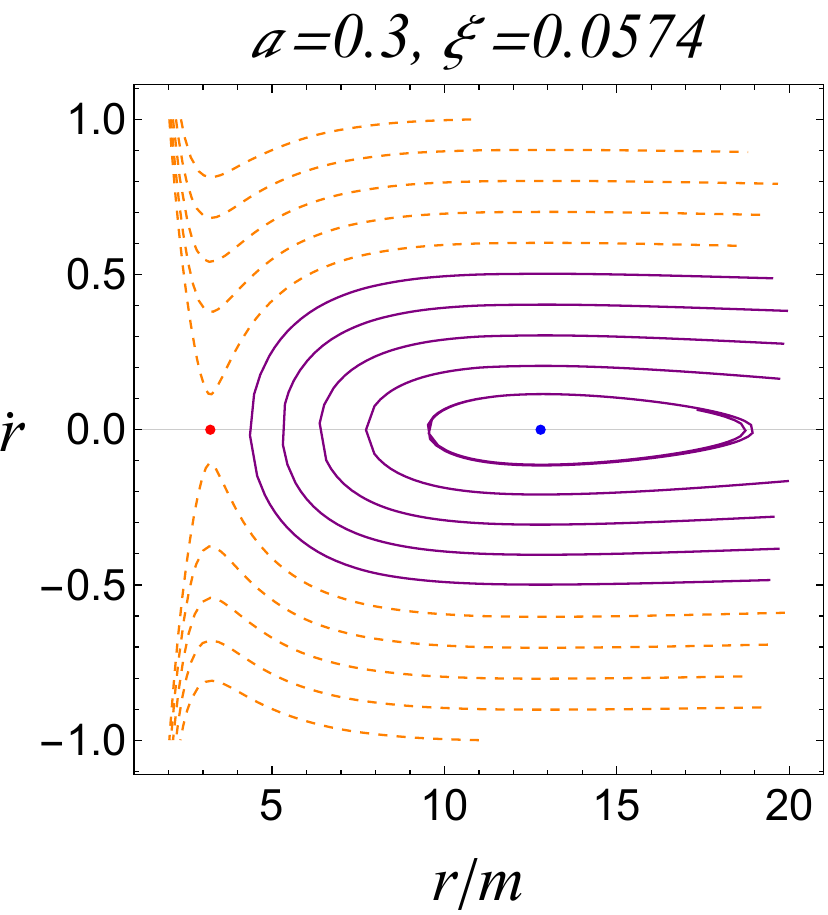}  

\caption{Phase diagrams in the   $r-\dot{r}$  plane for timelike geodesics in dCS Spacetime, shown for spin parameters from left to right $a=0.3$, and $\xi=0$, $\xi=0.0336$ and $\xi=0.0576$; The red highlighted point $x_1=(0,r_1)$ is a saddle point, nearby solutions  in the neighborhood are unstable (dashed); the blue highlighted critical point $x_2=(0,r_2)$ is a stable center, and solutions  near are stable closed orbits (solid). It is observed that the critical points exhibit a slight displacement along the $r$ axis as variations occur in the spin parameter $a$.}
\label{fig5}
\end{figure}

\begin{table}[htbp]
\caption{Range of stable circular orbits for timelike particles corresponding to various values of $a$ and $\xi$. Stable circular orbits are maintained at radii $r^{*}_{+}>6$, designating the innermost stable circular orbit, while unstable circular orbits occur at $r^{*}_{+} < 6$.}\label{tab5}%
\begin{tabular}{@{}lllll@{}}
\toprule
 \textbf{a} & \textbf{Range of stable circular orbits} & \textbf{\(\xi = 0.0336\)} & \textbf{\(\xi = 0.0574\)} \\ \hline
0 (Sch) & $r^{*}_{+} > 6$ & $r^{*}_{+} > 6$ & $r^{*}_{+} > 6$ \\ 
0.2 & $r^{*}_{+} > 5.33125$ & $r^{*}_{+} > 5.33141$ & $r^{*}_{+} >5.33152 $ \\
0.3 & $r^{*}_{+} > 4.9852$ & $r^{*}_{+} >4.98544 $ & $r^{*}_{+} >4.98561$ \\
\botrule
\end{tabular}

\end{table}

\subsection{KCC stability}
\begin{figure}[ht]
    \centering
    \includegraphics[width=0.48\linewidth]{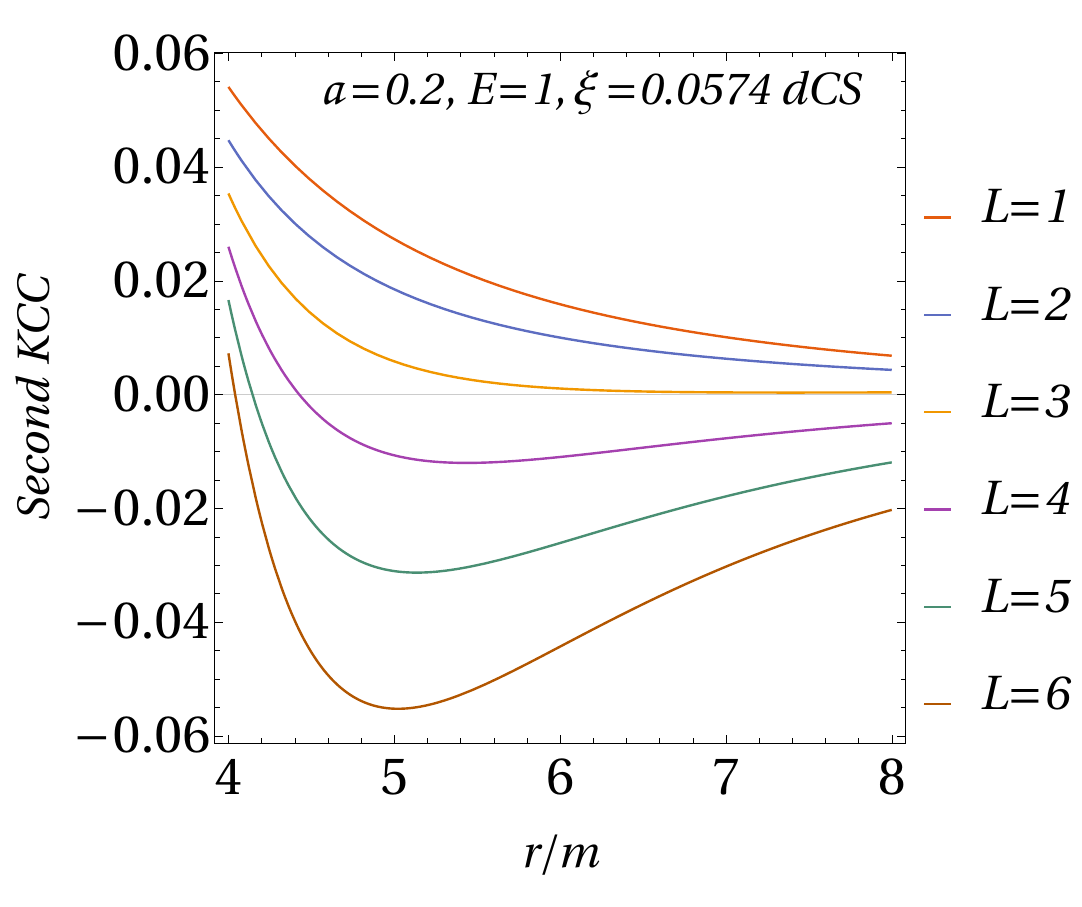}
\includegraphics[width=0.48\linewidth]{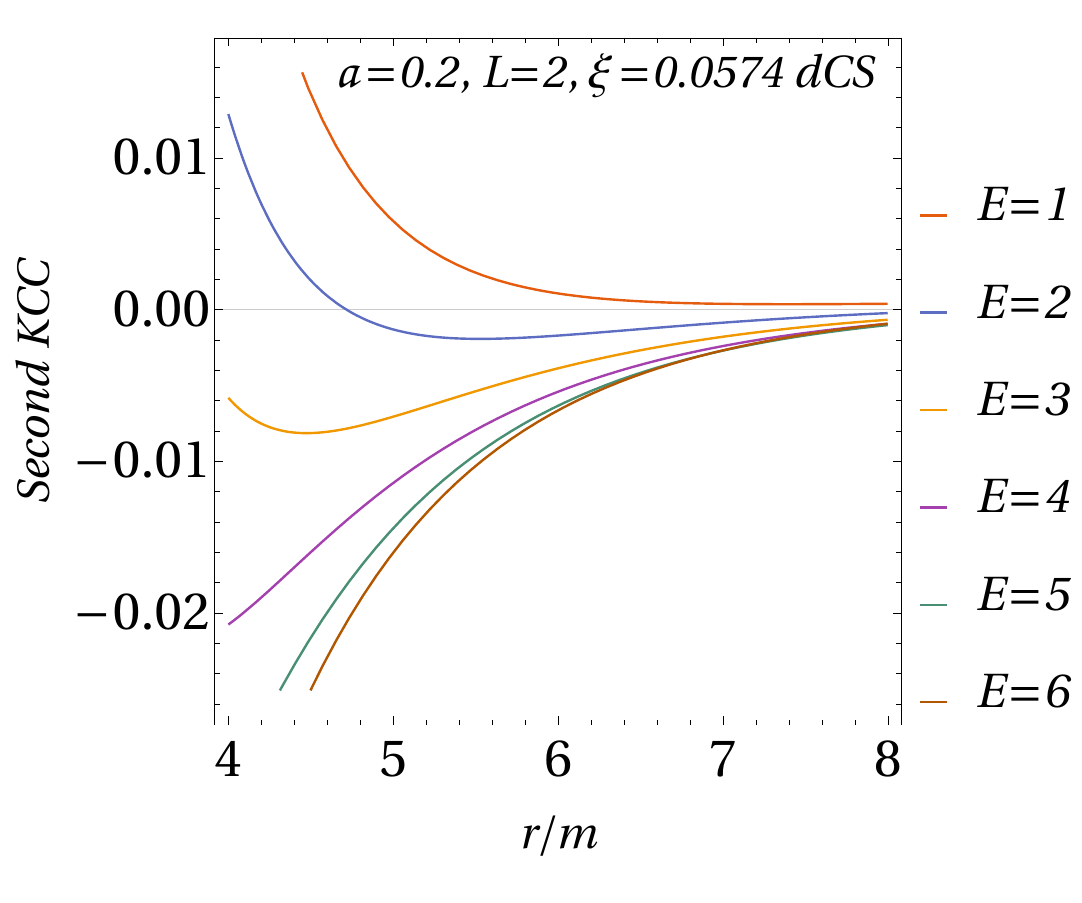}
     
    \caption{The figure on the left illustrates the second KCC $P^{1}_{1}$, maintaining constant parameters for spin $a$, energy $E$, and the coupling constant $\xi$ while varying the angular momentum $L$. Conversely, the figure on the right depicts the second KCC $P^{1}_{1}$, where the parameters for spin $a$, angular momentum $L$, and coupling constant $\xi$ are held constant as the energy $E$ is varied.}
    \label{fig:seconkccVSr}
\end{figure}

Our goal in this analysis is to establish the second KCC invariant, referred to as the deviation curvature tensor, in order to assess Jacobi stability. By employing the expression for the effective potential of dCS spacetime, denoted by $V'_{\text{Eff}}$, we derive
\begin{equation}
    \overset{..}{r} + V_{\text{Eff}}=0
\end{equation}
Contrasting the previous equation with the general second-order differential equation used in KCC theory
\begin{equation}
    \frac{d^2 x^i}{dt^2} + 2G^i(X,Y) = 0,
\end{equation}
it is evident that the second terms in both equations are identical, thus it can be inferred that
\begin{equation}
    G^{1}(r,p) =G^{1}(r,p)_\text{K}+\xi G^{1}(r,p)_\text{CS},
\end{equation}
with 
\begin{align}
    G^{1}(r,p)_\text{K}&=\frac{3 (-a E + L)^2 m + (a^2 (-1 + E^2) - L^2) r +  m r^2}{ r^4},\\
    G^{1}(r,p)_\text{CS}&= \frac{15 a E L r^3 + 5 a^2 m \left(-9 L^2 + \frac{E^2 \mathcal{F}^{*}_{12} r^3}{\mathcal{F}_{2}^2}\right)}{4 r^{10}}.
\end{align}
The derivatives of these functions are
    \begin{align}
         G'^{1}(r,p) &= G'^{1}(r,p)_\text{K}+\xi G'^{1}(r,p)_\text{CS},\\
        G'^{1}(r,p)_\text{K}&=\frac{24 a E L m + 3 L^2 \mathcal{F}_{4} - 2 m r^2 - 3 a^2 (-r + E^2 \mathcal{F}_{4})}{r^5}, \\
        G'^{1}(r,p)_\text{CS}&=-\frac{5 a \left(-90 a L        
    ^2 m \mathcal{F}_{2}^3 +21 E L \mathcal{F}_{2}^3 r^3 + 
8 a E^2 m r^3 \mathcal{F}^{**}_{7}\right) }{-4 r^{11}\mathcal{F}_{2}^3}.
    \end{align}
The nonlinear connection for this system is determined by $N_{1}^{1} = \frac{\partial G^1}{\partial p} = 0,$ and the Berwald connection is established as $G_{11}^1 = \frac{\partial N_{1}^{1}}{\partial p} = 0
$. 
The equation for the deviation curvature tensor \eqref{kcc. curvature tensor} takes the form
\begin{equation}
      P_1^1=-2\frac{\partial G^1}{\partial r}-2G^1G_{11}^1+p\frac{\partial N_1^1}{\partial r}+N_1^1N^1_1.
\end{equation}
Having derived all the ingredients, we are prepared to formulate the second KCC invariant as follows:
\begin{align}
    P_1^1&=P^1_\text{1 K}+\xi P^1_\text{1 CS},
\\
    P^1_\text{1 K}&=\frac{-48 a E L m + 4 m r^2 - 6 L^2 \mathcal{F}_{4} + 6 a^2 (-r - E^2 \mathcal{F}_{4})}{r^5},
\\
    P^1_\text{1 CS}&=\frac{5 a \left(-90 a L^2 m \mathcal{F}_{2}^3 +21 E L \mathcal{F}_{2}^3 r^3 + 
8 a E^2 m r^3 \mathcal{F}^{**}_{7}\right) }{-2 r^{11} \mathcal{F}_{2}^3}.
\end{align}
Evaluating the second invariant KCC at the critical point (0,$r_{*}$), the Jacobi stability is determined. The results are illustrated in Fig.~\ref{fig:seconkccVSr}.

The behavior of the Second KCC was analyzed as a function of the radial coordinate $r$ with varying angular momentum ($L$) and energy ($E$). The initial set of plots in Fig.~\ref{fig:seconkccVSr} evaluated the behavior of the second KCC characterized by $L$, with energy, spin parameter, and coupling constant maintained at fixed values, denoted as $E=1.0$, $a=0.2$, and $\xi=0.0574$ respectively. The analysis yielded the following insights: there was a shift in the stability regions at lower values of $L$, where the Second KCC predominantly displayed positive values throughout the domain of $r$, indicating an inherent instability. With the increase of $L$, the Second KCC started to present negative values over specific intervals marked by $r$, signifying the development of stable regions. For minimal values of $L$, the Second KCC similarly exhibited a predominance of positive values over the domain $r$, again implying instability. As $L$ ascended, the emergence of negative value manifestations of Second KCC within certain ranges of $r$ marked the formation of stable regions. The presence of higher angular momentum brought about a stabilizing effect, as it introduced additional terms that mitigate the destabilizing influences stemming from the radial and coupling-dependent components of the Second KCC. This results in an expanded region of $ r $ where the stability conditions delineated by the Second KCC  are fulfilled. During critical transitions, for median values of $L$, the Second KCC oscillated between stable and unstable regions across the domain, intersecting zero at critical radii. These transitions imply potential bifurcations or alterations in the nature of the trajectory dynamics.

The effects of energy $E$ on stability were examined in the second set of plots \ref{fig:seconkccVSr}, where the influence of varying $E$ was investigated while constants $L=3.0$, $a=0.2$, and $\xi=0.0574$ remained unchanged. At lower energy levels, the Second KCC reveals extensive regions of negative values, indicating stability across a broad spectrum of $r$. With an increase in $E$, the magnitude of the positive terms within the Second KCC escalates, resulting in a reduction of the stable region and a shift in the critical radius at which stability transitions occur. The compression of stability zones is observed for higher values of $E$, leading to a considerable reduction in stable regions and an increase in positivity across the majority of the domain, indicating heightened instability. This analysis underscores the destabilizing impact of energy, which enhances the influence of centrifugal and radial forces, thereby overpowering the stabilizing contributions from other terms. 
These findings suggest that systems characterized by moderate to high angular momentum and low energy are predisposed to stable trajectories, similar to the behavior observed in the Kerr case. Conversely, high-energy systems with low angular momentum are likely to exhibit instability across a broad range of $ r $.

Our results indicate that stability is not significantly impacted by variations in the coupling constant $\xi$. Instead, changes in $\xi$ primarily alter the location where the second KCC invariant intersects the $r$-axis, as shown in Figure~\ref{fig:seconkccVSr}. This behavior is observed as a lateral displacement of the critical points, depending on the value of $\xi$, without affecting the overall stability of the system. Such displacements can be interpreted as vertical shifts in the second KCC function, reflecting analogous behavior observed in parameters such as energy, spin, and angular momentum. These findings underscore that while $\xi$ modifies the geometric characteristics of the system, it does not inherently enhance or diminish its stability.

\section{Discussion and conclusions}
In this paper, we have studied the stability of time-like geodesics in the equatorial plane of the slowly rotating dCS black hole by analyzing the effect of the spin parameter $a$ and the coupling parameter $\xi$ on the effective potential. We used linear stability, KCC stability analysis, and phase portrait analysis. With the aim of aligning with realistic modified gravity scenarios, the coupling parameter $\xi$ is restricted with observational data of black hole shadows.  

From the analysis of Eq.~\eqref{potenciald}, stability analysis of critical points was carried out, allowing us to determine stable critical points and a range of stable orbits of the nonlinear system. {We find that, regardless of the values of the spin and CS coupling parameters --within the ranges used in this work -- the critical point $x_1=(0,r_1)$ is consistently Lyapunov unstable, while $x_2=(0,r_2)$ remains stable, and this behavior can also be observed in the phase diagrams. This indicates a robust saddle-point behavior for this configuration. }
Although the values of $V''(r_1)$ at the first critical point $x_1=(0,r_1)$ become increasingly negative as $a$ increases, indicating a worsening instability, the stability at $x_2(0,r_2)$ remains consistently positive, suggesting that perturbations will tend to stabilize the system at this point. The analysis reveals that the system consistently exhibits saddle-point behavior in different values of $a$ and $\xi$. Although $r_1$ is Lyapunov unstable in all scenarios, $r_2$ is still stable.

Furthermore, the analysis of the second KCC invariant reveals how spin and Chern-Simons correction influence the stability of black hole solutions. Although Schwarzschild black holes remain stable, increasing spin and introducing dCS corrections can lead to instability, as evidenced by the upward shifts and crossings of the second KCC invariant into positive values. The following summarizes the derived results:
\begin{itemize}
    \item Schwarzschild:  
    The system is stable, with the second KCC invariant consistently negative, showing strong Jacobi stability.

    \item Kerr
    :  Introducing a small spin reduces stability but does not necessarily lead to instability. With higher spin, the system becomes more unstable.
    
    \item dCS 
    : Increasing both the spin and the dCS corrections gradually makes the system less stable, with the second KCC invariant approaching zero or turning positive in some regions. 
    \item In the range of parameters analysed, the stability exhibits negligible sensitivity to changes in $\xi$.

\end{itemize}
Both methods indicate that $x_1(0,r_1)$ remains unstable in all scenarios analyzed. The increasing negativity of $V''(r_1)$ as $a$ increases aligns with the behavior of the second KCC invariant, which shows an increasing instability as the spin increases. The stability observed at $x_2(0,r_2)$ in the Lyapunov analysis is corroborated by the second KCC invariant remaining negative for Schwarzschild configurations. 
The transition from stability to instability as the parameters are varied is consistently represented in both analyses. For Kerr solutions with higher spin ($a = 0.3$), both methods indicate an increased instability: Lyapunov's analysis shows that $V''(r_1)$ becomes more negative, whereas the second KCC invariant signifies instability by moving upwards to positive values. 
This suggests that both methods are compatible and can be used together to offer a more comprehensive understanding of the stability characteristics of black hole solutions under varying conditions of spin and dCS corrections. Additionally, in the appropriate limits, the results obtained are in agreement with those reported by Ref.~\cite{Hossein,Singh_2022} for Schwarzschild and Kerr black holes.

We observe some benefits of using the KCC approach.
Provides a more comprehensive framework for understanding the robustness of solutions to second-order differential equations. Unlike classical stability methods, which may focus primarily on linearization and local behaviors near equilibrium, the KCC stability extends to the geometric and structural properties of the system. This simplifies the identification of how small changes in initial conditions or internal parameters gradually change the stability properties of the system, while also giving a more nuanced view of the system's global structure, beyond just equilibrium points. As a result, this method is well suited for systems where classical linear stability analyzes may fall short or provide insufficient insight into the long-term behavior of the system under perturbations.

Finally, regarding the astrophysical consequences of the CS modification, we find that, for timelike geodesics, the innermost stable circular orbit $r_{\text{isco}}$ delineates the boundary of the stable accretion disk region and plays an essential role in determining fundamental physical properties and electromagnetic signatures of the accretion disk, including accretion rate, radiant energy, observed luminosity, and temperature~\cite{Harko_2009,Harko_2010,Heydari_Fard_2021,heydarifard2024accretiondiskimagesrotating}. In our study, for values of $\zeta$ that are inspired by those found in~\cite{rodríguez2024shadowsblackholesdynamical}, stability regions are obtained that closely resemble those predicted by GR, the difference in $r_{\text{isco}}$ values due to variations in $\zeta$ are up to $0.008\%$. However, with significantly larger variations in $\zeta$, substantial differences arise, which impact $r_{\text{isco}}$ and lead to discernible changes in electromagnetic signatures, as indicated in~\cite{Harko_2009}.
For null geodesics, which are not explicitly analyzed in this work, the stability region is associated with the photon sphere. Variations in $\zeta$ cause deformations in this region, thereby altering the shadow shape, as investigated in~\cite{rodríguez2024shadowsblackholesdynamical, Meng_2023}.
These astrophysical predictions present opportunities for comparison with future observations from the Event Horizon Telescope (EHT).

\section{Acknowledgments}
The authors appreciate the support provided by CONAHCyT through grants 932448, 519369 and DCF-320821, which made this research possible.

\bibliography{refs}

\end{document}